\newcommand*{\AuthorVersionConf}{}

\ifdefined\AuthorVersionTOG
    \documentclass[acmtog,authorversion,balance=true]{acmart}
\else
    \ifdefined\AuthorVersionConf
        \documentclass[sigconf,authorversion,screen,balance=true]{acmart}
    \else
        \documentclass[sigconf,screen,balance=true]{acmart}
    \fi
\fi

\usepackage{stfloats}
\copyrightyear{2024}
\acmYear{2024}
\setcopyright{rightsretained}
\acmConference[SIGGRAPH Conference Papers '24]{Special Interest Group on Computer Graphics and Interactive Techniques Conference Conference Papers '24}{July 27-August 1, 2024}{Denver, CO, USA}
\acmBooktitle{Special Interest Group on Computer Graphics and Interactive Techniques Conference Conference Papers '24 (SIGGRAPH Conference Papers '24), July 27-August 1, 2024, Denver, CO, USA}
\acmDOI{10.1145/3641519.3657464}
\acmISBN{979-8-4007-0525-0/24/07}

\title{\nbvh: Neural ray queries with bounding volume hierarchies}

\author{Philippe Weier}
\email{weier@cg.uni-saarland.de}
\orcid{0000-0003-4276-1804}
\affiliation{%
    \institution{Saarland University}
    \country{Germany}
}

\author{Alexander Rath}
\email{rath@cg.uni-saarland.de}
\orcid{0000-0002-6084-2942}
\affiliation{%
    \institution{Saarland University}
    \country{Germany}
}

\author{Élie Michel}
\email{emichel@adobe.com}
\orcid{0000-0002-2147-3427}
\affiliation{%
    \institution{Adobe}
    \country{France}
}

\author{Iliyan Georgiev}
\email{igeorgiev@adobe.com}
\orcid{0000-0002-9655-2138}
\affiliation{%
    \institution{Adobe}
    \country{UK}
}

\author{Philipp Slusallek}
\email{slusallek@cg.uni-saarland.de}
\orcid{0000-0002-2189-2429}
\affiliation{%
    \institution{Saarland University}
    \country{Germany}
}

\author{Tamy Boubekeur}
\email{boubek@adobe.com}
\orcid{0000-0001-5985-0921}
\affiliation{%
    \institution{Adobe}
    \country{France}
}

\usepackage{wrapfig}
\usepackage{colortbl}

\ifdefined\TAPSVersion
\usepackage[capitalize,nameinlink]{cleveref}

\else
\usepackage[capitalize,nameinlink]{cleveref}
\fi

\usepackage{tabularx}
\usepackage{mathtools}
\newcommand\CIRCLE{\huge{\ensuremath{{\bullet}\mathllap{\circ}}}}

\newcommand{\nbvh}{$\mathcal{N}$-BVH\xspace}

\definecolor{RevisionColor}{RGB}{2,95,217}
\newcommand{\revision}[1]{#1}

\definecolor{PhilippeColor}{RGB}{2,95,217}
\definecolor{ElieColor}{RGB}{170,46,200}
\definecolor{TamyColor}{RGB}{20,180,22}
\definecolor{IliyanColor}{RGB}{210,20,22}
\definecolor{AlexColor}{RGB}{240,130,10}

\usepackage{xspace}

\newcommand{\GB}{\textsc{GB}}
\newcommand{\MB}{\textsc{MB}}

\begin{document}

%%%%%%%%%%%%%%%%%%%%%%%%%%%%%%%%%%%%%%%%%%%%%%%%%%%%%%%%%%%%
% Teaser figure
%%%%%%%%%%%%%%%%%%%%%%%%%%%%%%%%%%%%%%%%%%%%%%%%%%%%%%%%%%%%

\begin{teaserfigure}
    \centering
    \vspace{2mm}
    \includegraphics{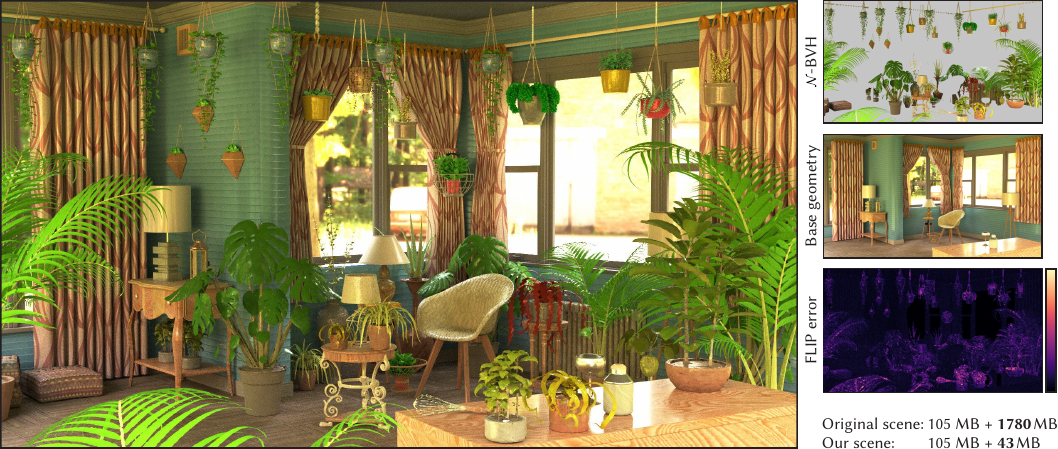}%
    \vspace{-1mm}
    \caption{
        \nbvh provides a compressed representation for ray queries against complex 3D assets, integrating seamlessly into standard ray-tracing pipelines. Here, the rendered image (left) combines responses from our neural model (top right) with classical BVH/triangle ray queries (middle right), providing a faithful approximation to the original scene. Our model achieves a compression rate of 42$\times$ for the subset of geometry it represents, and 13$\times$ \mbox{over the entire scene.}
    }
    \label{fig:teaser}%
    \vspace{3mm}
\end{teaserfigure}

%%%%%%%%%%%%%%%%%%%%%%%%%%%%%%%%%%%%%%%%%%%%%%%%%%%%%%%%%%%%
% Abstract
%%%%%%%%%%%%%%%%%%%%%%%%%%%%%%%%%%%%%%%%%%%%%%%%%%%%%%%%%%%%

\begin{abstract}
    Neural representations have shown spectacular ability to compress complex signals in a fraction of the raw data size. In 3D computer graphics, the bulk of a scene's memory usage is due to polygons and textures, making them ideal candidates for neural compression. Here, the main challenge lies in finding good trade-offs between efficient compression and cheap inference while minimizing training time. In the context of rendering, we adopt a ray-centric approach to this problem and devise \nbvh, a neural compression architecture designed to answer arbitrary ray queries in 3D. Our compact model is learned from the input geometry and substituted for it whenever a ray intersection is queried by a path-tracing engine. \revision{While prior neural compression methods have focused on point queries, ours proposes neural \emph{ray queries} that integrate seamlessly into standard ray-tracing pipelines.} At the core of our method, we employ an adaptive BVH-driven probing scheme to optimize the parameters of a multi-resolution hash grid, focusing its neural capacity on the sparse 3D occupancy swept by the original surfaces. As a result, our \nbvh can serve accurate ray queries from a representation that is more than an order of magnitude more compact, providing faithful approximations of visibility, depth, and appearance attributes. The flexibility of our method allows us to combine and overlap neural and non-neural entities within the same 3D scene and extends to appearance level of detail.
    \ifdefined\AuthorVersionConf
        \vspace{1mm}
    \fi
\end{abstract}

%%%%%%%%%%%%%%%%%%%%%%%%%%%%%%%%%%%%%%%%%%%%%%%%%%%%%%%%%%%%




\maketitle

%%%%%%%%%%%%%%%%%%%%%%%%%%%%%%%%%%%%%%%%%%%%%%%%%%%%%%%%%%%%
\section{Introduction}
%%%%%%%%%%%%%%%%%%%%%%%%%%%%%%%%%%%%%%%%%%%%%%%%%%%%%%%%%%%%

Physically based rendering engines rely on a single fundamental operator to simulate light transport in a 3D scene: ray tracing. For realistic scenes encompassing complex surface meshes
% geometries
% in the form of surface mesh 
(see \cref{fig:teaser}), ray queries are accelerated using a hierarchical data structure that allows skipping empty space efficiently when tracing the rays in search for an intersection. Nevertheless, the memory footprint of the scene and of this structure is often challenging to manage, given the limited space available on graphics processing units (GPU). Recently, neural methods have demonstrated impressive ability to compress data, in particular spatial samplings, yet have been mostly designed to serve point queries, such as evaluating a shape represented via its \emph{signed distance function}. In this paper, we propose a new neural representation for 3D scene models designed specifically for ray queries that blends naturally into a typical ray tracer. Our key observation is that any neural compression model can be optimized efficiently as long as it is trained on samples that live close to the signal of interest. In our case, 3D surfaces are the signal of interest and are commonly structured in a \emph{bounding volume hierarchy} (BVH) to speed up their intersection test. We take inspiration from this very standard setup and propose to optimize a state-of-the-art neural data structure by embedding it into such a BVH; we call our new structure a \emph{Neural BVH} or \nbvh.

At training time, we use the scene's BVH as a probing machine to generate ray queries/responses training pairs only close to the surfaces for our neural model to learn. At rendering time, our \nbvh inherits the natural empty-space skipping behavior of a standard BVH and serves neural ray queries, preventing full, deep BVH traversal and hence storage. Our \nbvh can be used concurrently with a standard BVH, to overlap neural and non-neural assets in a single scene. Its training takes only a few minutes even on large scenes and its runtime response includes visibility, depth, and appearance attributes for arbitrary rays. Additionally, \nbvh proposes a simple level-of-detail (LoD) scheme, by refining multiple error-driven cuts in its underlying tree structure and optimizing the neural model concurrently at all their nodes.

%%%%%%%%%%%%%%%%%%%%%%%%%%%%%%
\paragraph{Contributions}
%%%%%%%%%%%%%%%%%%%%%%%%%%%%%%

We introduce the following novel elements:
\begin{itemize}
    \item
          A new hybrid neural data structure, \nbvh, which encodes signals such as depth, normal, or appearance attributes so that they can be efficiently queried by a ray, and focuses its neural capacity on the sparse subset of 3D space spanned by surfaces;
    \item
          A neural ray-intersection query mechanism (\cref{sec:node_encoding}) which can serve any path tracer with faithful intersection approximations;
    \item
          A fast training scheme driven by a coarse-to-fine tree-cut optimization (\cref{sec:error_driven_tree_cut}), which automatically concentrates the training where the error is highest; this scheme provides a deep (resp.\ shallow) hierarchy where the geometry is hard (resp.\ easy) to learn and is oblivious to the actual tessellation of the input meshes, e.g., a densely subdivided plane is perceived as simple;
    \item
          The ability to jointly define multiple adaptive neural levels-of-detail (\cref{sec-bvh-lod}), specifying a target node count for each related tree cut.
\end{itemize}
Basing \nbvh on hash grids copes ideally with BVH empty-space skipping: typically, the vast majority of the backpropagation gradients are non-zero during training even if only a small amount of the scene's volume is actually sampled. This sparse adaptation translates into encoded points landing close to the geometry, inducing faster and easier learning at a reduced memory cost. This allows for a very lightweight hash grid encoding---in the order of a few megabytes---where we favor BVH depth over neural capacity growth.

%%%%%%%%%%%%%%%%%%%%%%%%%%%%%%
\paragraph{Application spectrum}
%%%%%%%%%%%%%%%%%%%%%%%%%%%%%%

We demonstrate the benefit of our \nbvh with two application scenarios: \emph{hybrid path tracing} (\cref{sec:hybrid_path_tracing}) which combines neural and non-neural assets in a single pipeline; and \emph{neural appearance prefiltering} (\cref{sec:neural_prefiltering}) which alleviates memory consumption and render times while providing equal or lower prefiltering error on large complex models appearance.

\revision{The source code of our implementation is publicly available at \url{https://github.com/WeiPhil/nbvh}.}
%%%%%%%%%%%%%%%%%%%%%%%%%%%%%%
\begin{figure*}[t]
    \begin{center}
        \includegraphics{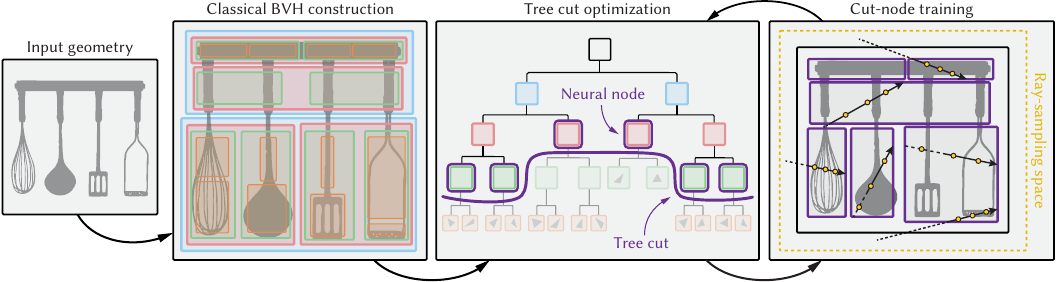}%
        \vspace{-2mm}
        \caption{
            Our lightweight \nbvh is a shallow hierarchy whose leaf nodes simply store bounds within which we search for ray intersections by querying a neural geometry representation. The leaves (in purple outlines) represent a cut in a classical BVH over the input geometry; we optimize the cut by iteratively splitting the leaves with largest inference error. After each splitting step we train our neural model within the cut-node bounds by sampling random rays in the scene.
        }
        \label{fig:pipeline}
    \end{center}
\end{figure*}
%%%%%%%%%%%%%%%%%%%%%%%%%%%%%%

%%%%%%%%%%%%%%%%%%%%%%%%%%%%%%%%%%%%%%%%%%%%%%%%%%%%%%%%%%%%
\section{Related work}
%%%%%%%%%%%%%%%%%%%%%%%%%%%%%%%%%%%%%%%%%%%%%%%%%%%%%%%%%%%%

%%%%%%%%%%%%%%%%%%%%%%%%%%%%%%Z
\paragraph{Neural implicit representations}
%%%%%%%%%%%%%%%%%%%%%%%%%%%%%%

Neural radiance fields (NeRF) \cite{mildenhall2020} and their recent adaptations to interactive and real-time graphics~\cite{Mueller22} use implicit neural representations forming neural fields~\cite{xie2022nf} for novel view synthesis. Neural representations' adaptability, coupled with sparse~\cite{Mueller22} and compressed~\cite{takikawa2022} coordinate encodings, efficiently store high-dimensional functions, representing high-frequency content in spatial and angular domains. The necessity for accelerated coordinate-based representation has mainly been explored in NeRF-related techniques where the dense sampling of a volume using ray-marching can lead to a prohibitively high number of network inferences. The (shallow) multi-layer perceptron (MLP) involved in the process is responsible for a large part of the inference cost which \citet{hedman2021snerg} proposes to avoid by accumulating features along rays to offload the MLP. \revision{\citet{yang2023tinc} also showed that a tree-structured MLP can further improve compression fidelity.} Recently, \citet{adaptiveshells2023} showed that further speed and accuracy can be gained with an adequate empty-space skipping strategy.
% , which they achieve by intersecting the inner and outer shells extracted from a learned SDF representation. 
Not only does this significantly reduce the number of network queries at runtime, but it can drastically improve the reconstructed signal as sampling is increased near the high-frequency content of the scene. Our approach is similar in spirit and relies on traditional scene acceleration data structures to provide an efficient empty-space skipping strategy. However, we are not restricted to NeRF applications as we aim at learning general ray-intersection queries which are at the core of many rendering applications.

%%%%%%%%%%%%%%%%%%%%%%%%%%%%%%
\paragraph{Geometric simplification}
%%%%%%%%%%%%%%%%%%%%%%%%%%%%%%

Silhouette and shape-preserving decimation \cite{Hoppe1996,garland97, Kobbelt1998} are widely used, often with artist supervision, in production environments. However, they can fail to accurately preserve the appearance of the original asset when high compression rates are demanded or the underlying geometry does not exhibit the geometric characteristic the algorithms are tailored to~\cite{Rossignac:1993:MR3,Lindstrom:2000:OOC,Schaefer:2003:AVC}. Point-based \cite{Pauly:2001:ESP} or statistical simplification \cite{Cook07} supports less structured shapes, and hybrid volume/surface approaches \cite{Gobbetti2005,Loubet2015} have been proposed but they usually focus on prefiltering rather than reducing the memory footprint of the scene.

%%%%%%%%%%%%%%%%%%%%%%%%%%%%%%
\paragraph{Neural compressed geometric reconstruction}
%%%%%%%%%%%%%%%%%%%%%%%%%%%%%%

A large body of work has shifted toward representing geometry as a signed distance field (SDF) rather than a more typical triangle representation. \citet{Park_2019_CVPR} were among the first to represent SDFs as deep neural networks, though were able to learn only low-frequency signals and at a high inference cost. Later, \citet{Takikawa2021} demonstrated how coordinate-based networks, implemented as an octree feature volume, allow much higher frequency SDFs to be reconstructed while also providing continuous levels of detail at a low memory footprint. As an alternative geometric representation, ACORN \cite{acorn_2021} proposes another type of coordinate-based network for learning a highly compressed and accurate occupancy field represented by an optimized space partitioning. While providing accurate reconstruction, it is specifically designed to answer point queries, which prevents its application to many rendering domains. The direct reconstruction of compressed triangle-based representation is a challenging problem even for coordinate-based networks as the underlying data is inherently discrete. \citet{neurallod2023} propose to compress and represent the prefiltered appearance while taking into account the correlation arising in structured geometry using a sparse voxel grid paired with a hash grid for effective compression of the learned signal. \citet{Feng22} tackle the problem of learning visibility and depth for arbitrary ray queries by proposing a ray-foot parameterization that prevents aliasing of different rays sharing the same intersection, thereby improving the learning process. However, their approach does not compress the geometric representation as it requires two large MLPs for inference, making it impractical for real-time use. Furthermore, the ability of their method to reconstruct complex high-frequency geometric signals is limited by the network's capacity. NeuralVDB \cite{kim2022neuralvdb} compresses the volume of a 3D scene with a set of overlapping domains, each equipped with an MLP mapping local voxel coordinates to voxel data. Its hierarchy is shallow and wide, and the method offers a significant compression ratio for point-queried volume data compared to previous iterations of the VDB representation.

%%%%%%%%%%%%%%%%%%%%%%%%%%%%%%
\paragraph{Hybrid neural path tracing}
%%%%%%%%%%%%%%%%%%%%%%%%%%%%%%

% The work of \cite{fujieda2023} focuses on optimizing ray-tracing performance through the learning of visibility. While their representation achieves real-time inference with a low memory footprint, it is constrained to fully static scenes and shadow rays only. Furthermore, their model is overfitted to both camera and light positions, preventing any type of interactivity or dynamic content.

The work of \cite{fujieda2023} focuses on optimizing ray-tracing performance through the learning of visibility. \revision{While their representation achieves real-time inference with a low memory footprint, their model only handles shadow ray queries and overfits to both camera and light positions, preventing any type of relighting or interactivity.}

%%%%%%%%%%%%%%%%%%%%%%%%%%%%%%%%%%%%%%%%%%%%%%%%%%%%%%%%%%%%
\section{Method overview}
%%%%%%%%%%%%%%%%%%%%%%%%%%%%%%%%%%%%%%%%%%%%%%%%%%%%%%%%%%%%

We propose a compact learnt representation for 3D surfaces with support for efficient ray-intersection queries (see \cref{fig:pipeline}). This representation replaces the input geometry with a coordinate-based neural model that we sample along rays to infer intersection point, surface normal, and appearance. Our model is based on a multi-resolution hash grid~\cite{Mueller22} which excels at compactly representing sparse but complex 3D signals.

Neural ray inference is subject to inherent reconstruction error. Beyond the capacity of the neural model, the parameterization of its input strongly influences its performance. Consequently, probing the model at sample points close to the geometry helps it to efficiently store intersection information. Dense sampling along the ray achieves high accuracy but is prohibitively costly; the main challenge is thus to sample the model sparsely but close to the 3D surface. To that end, we take inspiration from classical ray-intersection acceleration to cluster geometry in bounding boxes and probe the model only within the ray-box intersection intervals. Organizing these boxes into a bounding volume hierarchy, dubbed \nbvh, recovers the benefits of traditional acceleration structures, namely (approximate) front-to-back probing for early termination and efficient empty-space skipping.

Note that we use a single global neural model encompassing the entire geometry, with a BVH to provide efficient training and ray inference in the near-field of the model. Since no geometrical primitives, no textures, and only a shallow BVH need to be stored, our method achieves significant compression rates.

We begin by describing our representation and the practical need for its adaptive probing (\cref{sec:node_encoding}). We then present our \nbvh structure and its error-driven construction, which we extend to support level-of-detail and further accelerate inference (\cref{sec:neural_bounding_volume_hierarchy}).

%%%%%%%%%%%%%%%%%%%%%%%%%%%%%%%%%%%%%%%%%%%%%%%%%%%%%%%%%%%%
\section{Neural ray query}
\label{sec:node_encoding}
%%%%%%%%%%%%%%%%%%%%%%%%%%%%%%%%%%%%%%%%%%%%%%%%%%%%%%%%%%%%

Given a set of geometric elements and their bounding box, we train a neural model to answer intersection queries for rays that intersect the box. Since our queries are low-dimensional, we adopt the current state-of-the-art approach which consists in combining spatial features with a small fully connected decoding module. Below we detail the encoding of our queries, our neural model, as well as its inference and training pipelines.

%%%%%%%%%%%%%%%%%%%%%%%%%%%%%%
\paragraph{Motivation}
%%%%%%%%%%%%%%%%%%%%%%%%%%%%%%

The simplest way to parameterize our query is via the ray-box entry and exit points; that is, to learn features on the bounding box surface. Unfortunately, this encoding exhibits poor correlation with the signal being learnt, i.e., the ray-intersection point. The intuition is that it stores information far away from the signal, which causes blur and loss in accuracy. We illustrate this behavior in the inline figure, where we focus on the yellow ray entry
    {% Begin group for wrap fig with configured margins
        \setlength{\columnsep}{4mm}%
        \setlength{\intextsep}{4pt}%
        \begin{wrapfigure}{r}{0pt}%
            \includegraphics{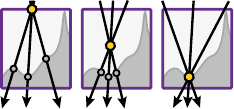}
        \end{wrapfigure}%
        point (the argument for the exit point is analogous). On the left, we see that rays going through this point have vastly different intersections, and the model is forced to aggregate (i.e., average) information across all of them at the (yellow) encoding point. Moving the encoding point inwards (middle subfigure) brings it closer to the surface where the ray-intersection points correlate more with one another. The ideal encoding location is then the intersection itself where we need to learn information only about that one point.

    }% End group for wrapfig. Do not remove the empty line above!

%%%%%%%%%%%%%%%%%%%%%%%%%%%%%%
\begin{figure}[!t]
    \begin{center}
        \includegraphics[page=1]{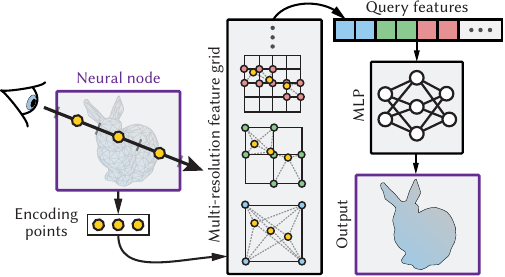}
        \caption{
            Our neural ray query pipeline. We sample uniformly along a given ray-box intersection interval and at each point collect features from a multi-resolution hash grid. The concatenated features are fed to an MLP to obtain a reconstructed signal (visibility, intersection, normal, appearance).
        }
        \label{fig:node_encoding}
    \end{center}
\end{figure}
%%%%%%%%%%%%%%%%%%%%%%%%%%%%%%

%%%%%%%%%%%%%%%%%%%%%%%%%%%%%%
\paragraph{Encoding, inference \& training}
\label{sec:training}
%%%%%%%%%%%%%%%%%%%%%%%%%%%%%%

In practice we obviously do not know the intersection location---it is what we seek to compute; we therefore resort to sampling the ray-box intersection interval at several locations which parameterize our query. We collect a set of features for each and decode the vector of concatenated features via a small multi-layer perceptron (MLP) to obtain the intersection response. \revision{The concatenation \emph{order} here is critical as it encodes the ray's direction}. The features are stored in a spatial multi-resolution hash grid~\cite{Mueller22}. We illustrate this inference scheme in \cref{fig:node_encoding}. We train the model by sampling the space of potential rays that intersect a node. For each ray, we (i)~sample its origin uniformly within the 50\%-inflated scene bounding box, (ii)~sample a uniform direction, and (iii)~query our model. The output is compared to the ground truth---obtained by intersecting the actual geometry, and the measured loss is back-propagated to the model's learnable parameters which are optimized via gradient descent. The loss depends on the specific signal being inferred---visibility, intersection, normal, etc.; we discuss the losses we use in \cref{sec:training_loss}.

%%%%%%%%%%%%%%%%%%%%%%%%%%%%%%
\paragraph{Sampling \& reconstruction quality}
%%%%%%%%%%%%%%%%%%%%%%%%%%%%%%

\begin{figure}[!t]
    \includegraphics[width=\linewidth]{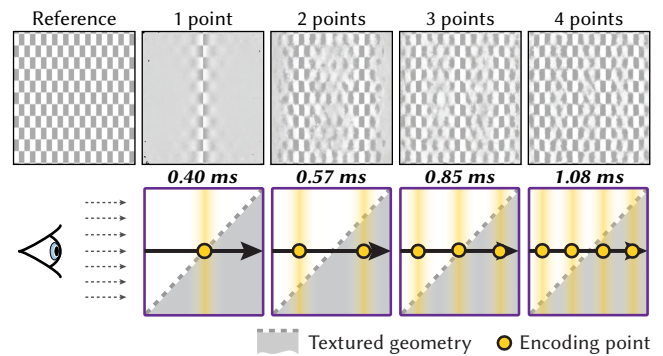}
    \caption{
        A surface textured with a square checkerboard, observed orthographically at a 45$^{\circ}$ angle. Reconstruction quality is high when our neural representation is probed close to the surface. Increasing the sampling rate along rays reduces error, but at proportionately higher inference cost.
    }
    \label{fig:node_encoding_experiment}
\end{figure}

Our ray encoding is obtained via stratified point-sampling along the ray (see \cref{fig:node_encoding}, left). We observe that to obtain accurate reconstruction it suffices that one of these points lies close to the surface, as the MLP learns to extract the relevant features from the concatenated vector. To maintain good accuracy we thus need to probe the model close to the surface. A naive way to achieve this is to increase the sampling rate along the ray, as we demonstrate in \cref{fig:node_encoding_experiment}. Unfortunately, denser sampling also increases inference time, to query the features from the grid and process them. Our \nbvh structure addresses this challenge.

%%%%%%%%%%%%%%%%%%%%%%%%%%%%%%
%%%%%%%%%%%%%%%%%%%%%%%%%%%%%%

%%%%%%%%%%%%%%%%%%%%%%%%%%%%%%%%%%%%%%%%%%%%%%%%%%%%%%%%%%%%
% \section{Neural bounding volume hierarchy (\nbvh)}
\section{Neural bounding volume hierarchy}
\label{sec:neural_bounding_volume_hierarchy}
%%%%%%%%%%%%%%%%%%%%%%%%%%%%%%%%%%%%%%%%%%%%%%%%%%%%%%%%%%%%

Scaling up to large and complex scenes calls for a sampling scheme that has the ability to sample sparsely but close to the geometry. To this end, we draw from decades of ray-tracing acceleration research: We can avoid intersection queries along ray intervals known to be traversing empty space. We split the input geometry into smaller, simpler pieces, enclosing each in a tight bounding box. Probing/training our model then need only consider ray segments inside these smaller boxes. Organizing the boxes into a bounding volume hierarchy (BVH) achieves our goals: it provides front-to-back ray traversal, empty-space skipping, and probing the model closer to the geometry. This lightweight structure is contained within the bounds of our neural model. We call it \emph{neural BVH}, or \emph{\nbvh}.

\begin{figure}[t]
    \includegraphics[width=1.03\linewidth]{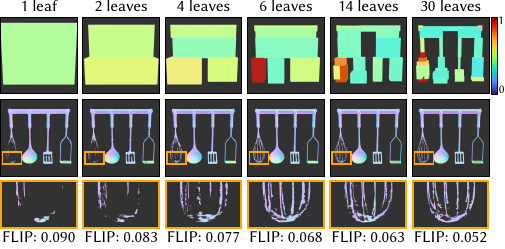}
    \caption{
        Increasing the number of \nbvh leaf nodes consistently improves the reconstruction quality. It also improves performance on this scene with low depth complexity where at most one neural inference per ray is needed. The top row visualizes the average training loss per node in false color.
    }
    \label{fig:increasing_leaf_node_count}
\end{figure}

Inference for a given ray proceeds by traversing the \nbvh and querying the neural representation upon reaching a leaf node as in \cref{fig:node_encoding}. The number of neural queries per ray is thus equal to the number of leaf nodes encountered before finding an intersection. In \cref{fig:increasing_leaf_node_count} we demonstrate that a shallow \nbvh already drastically improves the reconstruction compared to using a single neural node.

We illustrate the structure of our bounding hierarchy in \cref{fig:pipeline}. It resembles that of a classical BVH---the difference is in the depth and the leaf nodes. A classical BVH is usually deep, and its leaves contain geometric primitives. In contrast, our \nbvh is a very lightweight structure: it is shallow (1--2 orders of magnitude fewer nodes), and its leaves are hollow bounding boxes over larger geometric clusters that are represented implicitly by our neural ray-query model. That model dominates the memory footprint of our combined representation. We next describe how we construct our \nbvh.

%%%%%%%%%%%%%%%%%%%%%%%%%%%%%%
%%%%%%%%%%%%%%%%%%%%%%%%%%%%%%

%%%%%%%%%%%%%%%%%%%%%%%%%%%%%%%%%%%%%%%%%%%%%%%%%%%%%%%%%%%%
\subsection{Error-driven construction}
\label{sec:error_driven_tree_cut}
%%%%%%%%%%%%%%%%%%%%%%%%%%%%%%%%%%%%%%%%%%%%%%%%%%%%%%%%%%%%

We want our \nbvh to yield uniform inference error throughout the scene, i.e.,\ its leaves' error to be roughly equal. We adopt a top-down construction approach which alternates between model training and node splitting.

%%%%%%%%%%%%%%%%%%%%%%%%%%%%%%
\paragraph{Base-BVH cut optimization}
%%%%%%%%%%%%%%%%%%%%%%%%%%%%%%

Instead of building a bounding volume hierarchy from scratch, we leverage the structure of the readily available input-geometry BVH, or \emph{base BVH}. This relieves us from requiring explicit access to the geometry and reduces our task to finding a cut in this BVH. We optimize the cut iteratively starting from the root. In each step, we first train our neural model within the bounds of the nodes along the current cut for a number of iterations (detailed in \cref{sec:training_loss}). We then expand the cut by splitting the nodes with largest error, replacing each with its two children. The number of training iterations between consecutive cut-expansion operations increases progressively by a user-specified factor, and so does the number of node splits (as the tree grows bigger, and the cut longer). We terminate when a target node count is reached. The nodes along the base-BVH cut then become the leaf nodes of our \nbvh for which we run one final, longer round of training.

%%%%%%%%%%%%%%%%%%%%%%%%%%%%%%
\paragraph{Node error}
\label{sec:node_error}
%%%%%%%%%%%%%%%%%%%%%%%%%%%%%%

The error introduced by approximating a region in space (i.e., a node's content) with our representation is the product $q \cdot p$ of the node's training loss $q$ and the probability $p$ of a random ray hitting it. The loss we discuss in \cref{sec:training_loss} below; the probability is proportional to the node's surface area, though we estimate it by the fraction of training rays that hit the node, which gives us the flexibility to adjust the ray distribution (see \cref{sec:training_loss}). In our tests, ranking the nodes by the raw $q \cdot p$ product caused overly aggressive splitting of large and/or high-loss nodes, leading to unbalanced deep trees and low performance. We therefore use a ranking heuristic that dampens the error logarithmically: $r = 2 \log q + \log p$, where the additional factor 2 puts more weight on the loss to discourage splitting large, low-loss nodes too often.

\Cref{fig:increasing_leaf_node_count} shows a few \nbvh trees built using the ranking criterion $r$ to select the nodes to be split. In the supplemental document, we confirm on a complex mesh that fixed-depth, uniform node splitting yields worse reconstruction.

%%%%%%%%%%%%%%%%%%%%%%%%%%%%%%%%%%%%%%%%%%%%%%%%%%%%%%%%%%%%
\subsection{Node training and losses}
\label{sec:training_loss}
%%%%%%%%%%%%%%%%%%%%%%%%%%%%%%%%%%%%%%%%%%%%%%%%%%%%%%%%%%%%

To train our neural model, we sample rays uniformly inside the (inflated) root node, as described in \cref{sec:training}, and intersect them against the nodes in the current base-BVH cut. We need to train the neural model only within cut-node bounding boxes. Below we describe how we adjust the training to focus on high-error nodes and how we learn the different inference signals.

%%%%%%%%%%%%%%%%%%%%%%%%%%%%%%
\paragraph{Ray distribution}
%%%%%%%%%%%%%%%%%%%%%%%%%%%%%%

The goal of our \nbvh construction is to achieve spatially uniform inference error. Besides node splitting, we employ two techniques to make better use of our finite training budget. First, for each ray, we train only the first leaf node it intersects, which focuses the effort on more visible nodes. The intersected node is trained only with probability $\max(r / r_\mathrm{max}, 0.005)$, where $r$ is that node's error (\cref{sec:node_error}) and $r_\mathrm{max}$ is the largest error in the cut.

%%%%%%%%%%%%%%%%%%%%%%%%%%%%%%
\paragraph{Visibility}
%%%%%%%%%%%%%%%%%%%%%%%%%%%%%%

The visibility along a ray (segment) can be defined as a binary classification problem. We use a sigmoid activation, thresholding the output against 0.5, and a binary cross-entropy loss which shows better convergence than a typical $L_2$ loss \cite{simard2003}. Visibility (i.e., presence of intersection) is the easiest signal to learn, and we use it during training as an indicator whether intersection information (depth, normal, etc.) should be learnt for each ray segment. If the ground-truth visibility is 1 (i.e., no intersection), we set the losses for all other data to zero. This ensures that we do not learn unnecessary intersection information.

%%%%%%%%%%%%%%%%%%%%%%%%%%%%%%
\paragraph{Intersection point}
%%%%%%%%%%%%%%%%%%%%%%%%%%%%%%

For the intersection location, we have the
    {% Begin group for wrap fig with configured margins
        \setlength{\columnsep}{2.5mm}%
        \setlength{\intextsep}{3pt}%
        \begin{wrapfigure}{r}{0pt}%
            \small
            \setlength{\tabcolsep}{1pt}
            \begin{tabular}{ccccc}
                 &   & \textsf{Position} &   & \textsf{Distance} \\
                \rotatebox{90}{\textsf{\hspace{3.5mm}Global}}
                 &   &
                \cellcolor{black!80}
                \includegraphics[width=0.091\columnwidth,trim={263mm 40mm 263mm 20mm},clip]{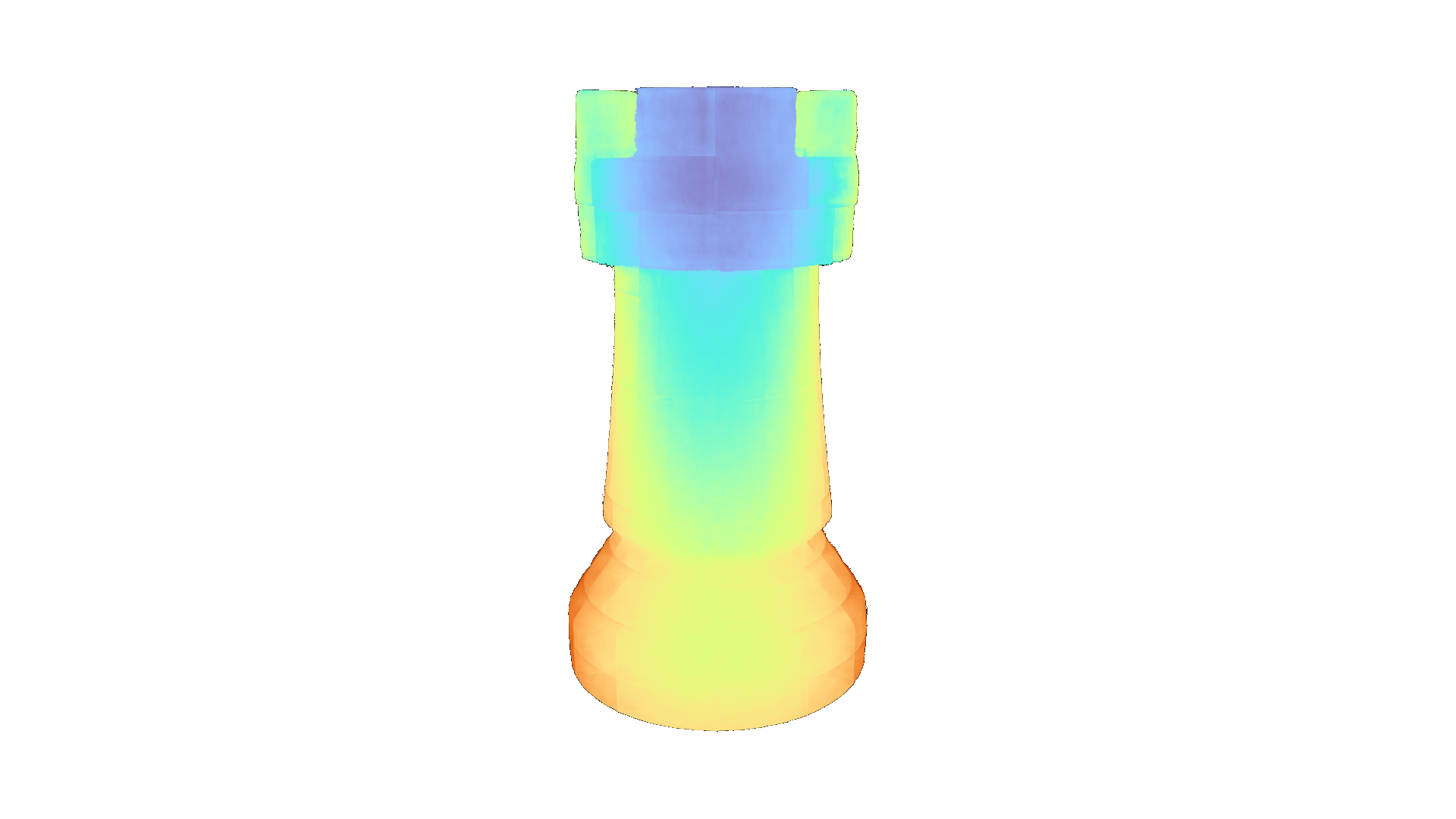}
                 &
                 &
                \cellcolor{black!80}
                \includegraphics[width=0.091\columnwidth,trim={263mm 40mm 263mm 20mm},clip]{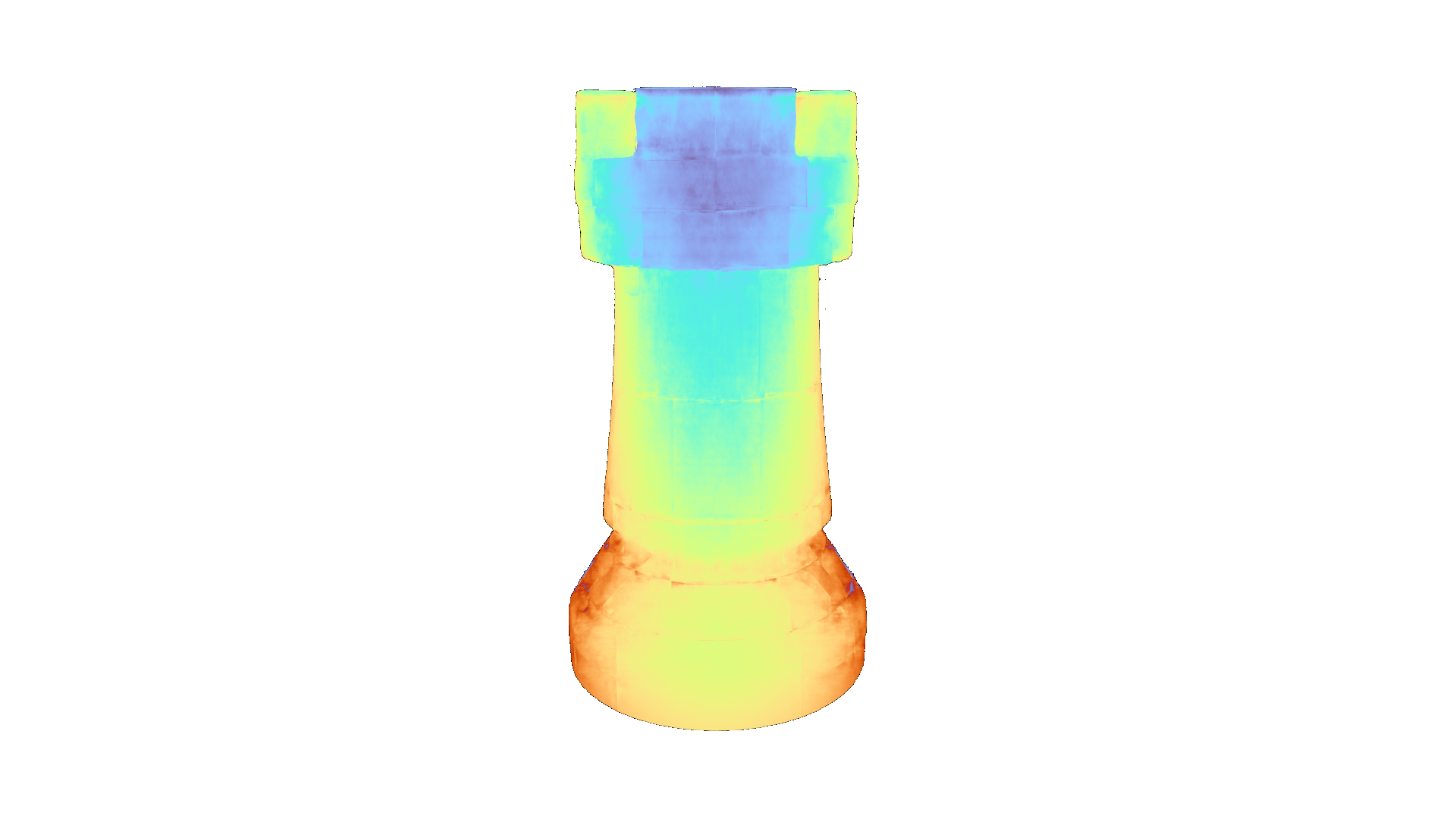}
                \\[-2.8mm]
                 &   &                   &
                \\
                \rotatebox{90}{\textsf{\hspace{4.5mm}Local}}
                 &   &
                \cellcolor{black!80}
                \includegraphics[width=0.091\columnwidth,trim={263mm 40mm 263mm 20mm},clip]{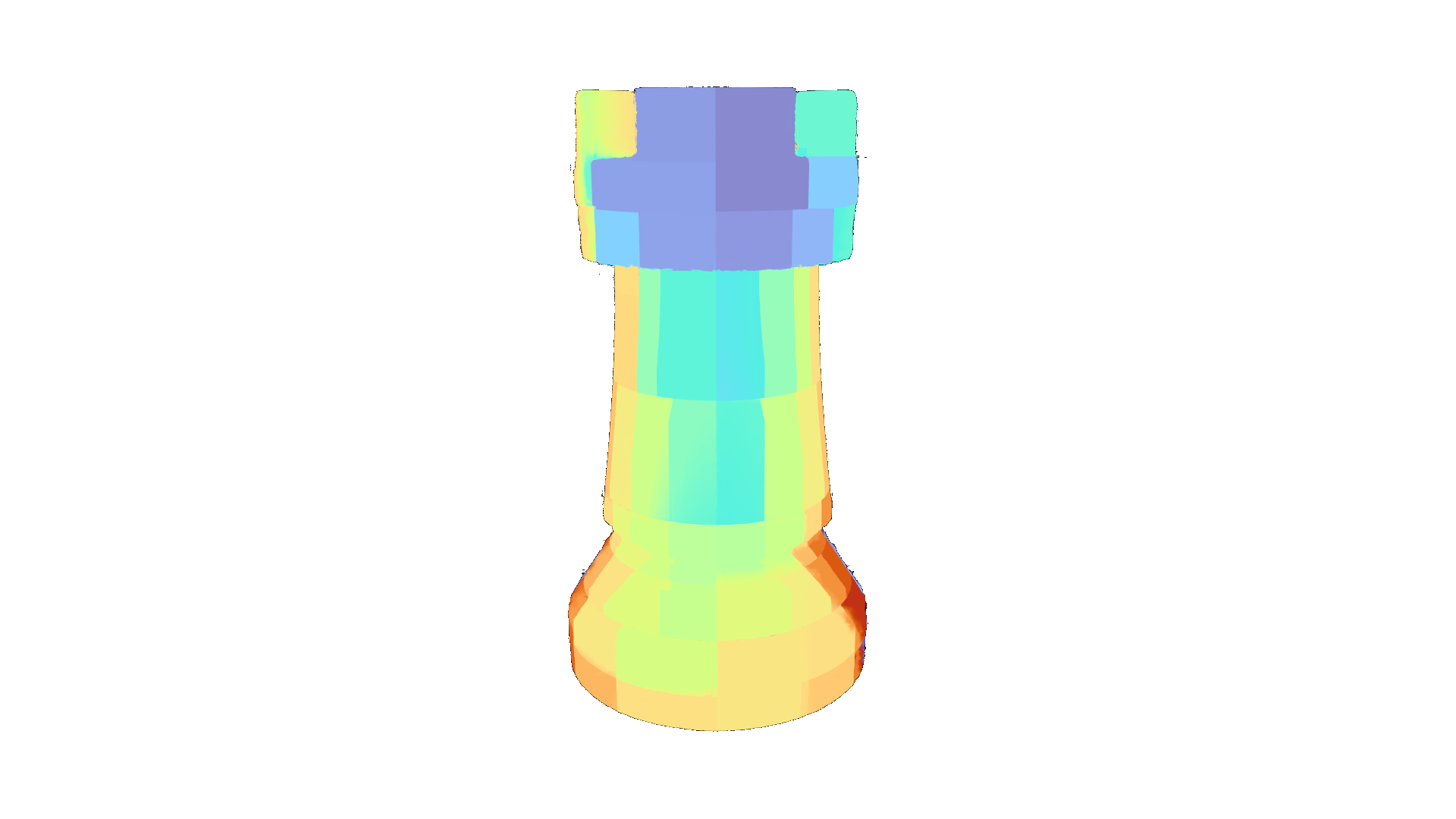}
                 &
                 &
                \cellcolor{black!80}
                \includegraphics[width=0.091\columnwidth,trim={263mm 40mm 263mm 20mm},clip]{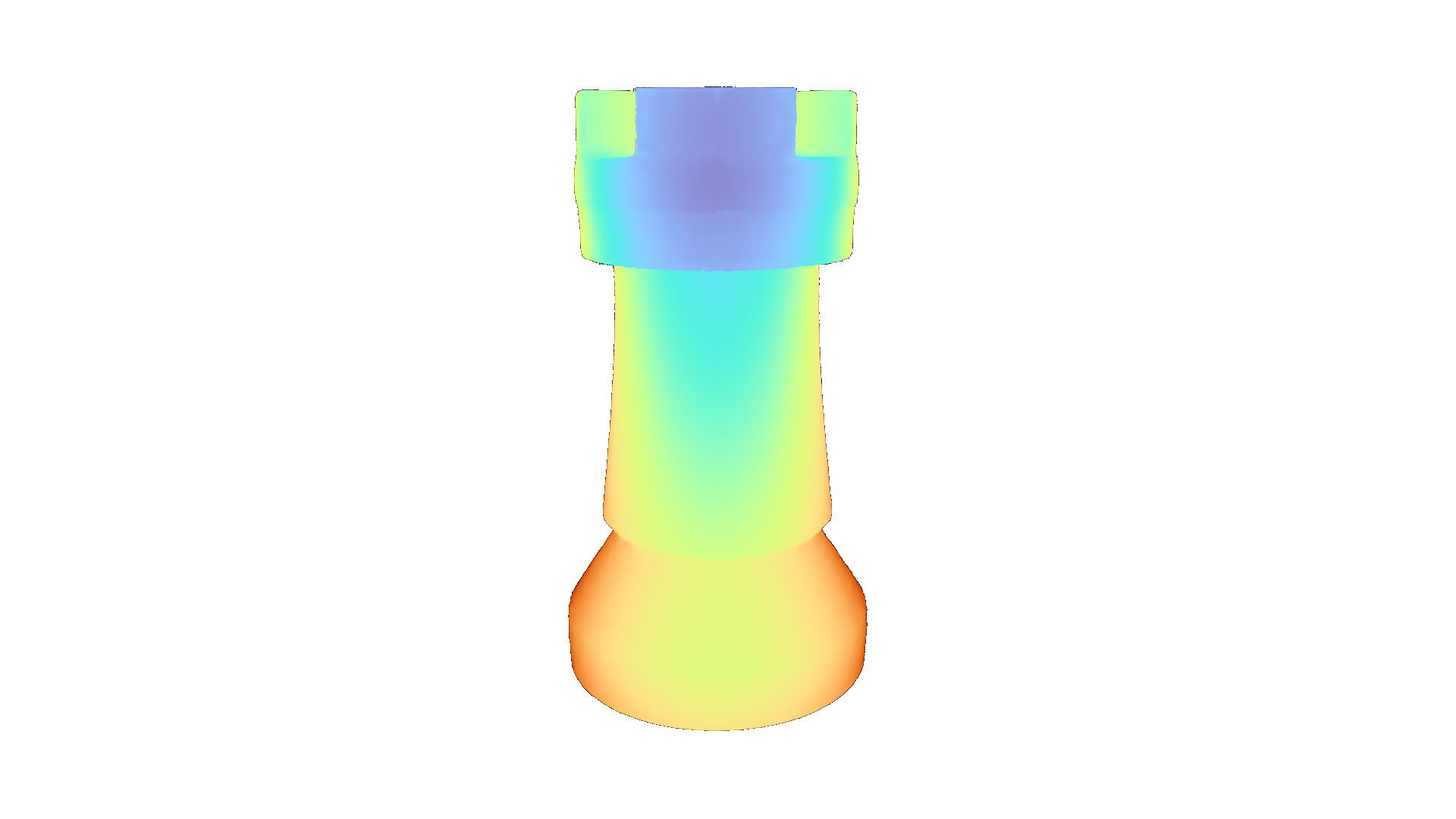}
            \end{tabular}
        \end{wrapfigure}%
        choice between learning the 1D distance along the ray or the 3D location directly. Furthermore, each of these can be learned locally, i.e., relative to the node's extent, or globally. We compare these four options in the inline figure. We find that locally learning the distance, with an $L_1$ loss, consistently achieves the highest quality. We attribute this behavior to the fact that a 1D signal is easier to learn/represent than a 3D one; constraining the signal to the unit interval also reduces its variation.

    }% End group for wrapfig. Do not remove the empty line above!

%%%%%%%%%%%%%%%%%%%%%%%%%%%%%%
\paragraph{Auxiliary intersection data}
%%%%%%%%%%%%%%%%%%%%%%%%%%%%%%

In our applications, the auxiliary intersection information that we learn are normal and albedo (the other BSDF parameters are fixed by the user). We use a relative $L_2$ loss for the albedo, which we found to perform better than a regular $L_2$ loss, even though the albedo signal varies between 0 and 1. For the normal, we use an $L_1$ loss.

%%%%%%%%%%%%%%%%%%%%%%%%%%%%%%
\paragraph{Combined loss}
%%%%%%%%%%%%%%%%%%%%%%%%%%%%%%

When using our representation in a standard rendering pipeline, e.g., our hybrid path-tracing application, we need to infer all four signals discussed above. The combined training loss we use is $L = 2 L_\mathrm{visibility} + 2 L_\mathrm{distance} + L_\mathrm{normal} + L_\mathrm{albedo}$, with more weight to visibility and distance for better geometric reconstruction.

%%%%%%%%%%%%%%%%%%%%%%%%%%%%%%%%%%%%%%%%%%%%%%%%%%%%%%%%%%%%
\subsection{Level of detail}
\label{sec-bvh-lod}
%%%%%%%%%%%%%%%%%%%%%%%%%%%%%%%%%%%%%%%%%%%%%%%%%%%%%%%%%%%%

Our representation can be trained simultaneously at different scales, or levels of detail (LoD). We apply a simple top-down approach to define \emph{multiple} base-BVH cuts during \nbvh construction, each defining an LoD. Since our split scheduling increases the number of tree-cut nodes exponentially, we register a new LoD at regular training iteration intervals. This results in a roughly linear increase in \nbvh depth between LoDs. During training, we randomly select a tree cut and train its nodes.

%%%%%%%%%%%%%%%%%%%%%%%%%%%%%%%%%%%%%%%%%%%%%%%%%%%%%%%%%%%%
\begin{figure*}
    \begin{center}
        \includegraphics[width=\linewidth]{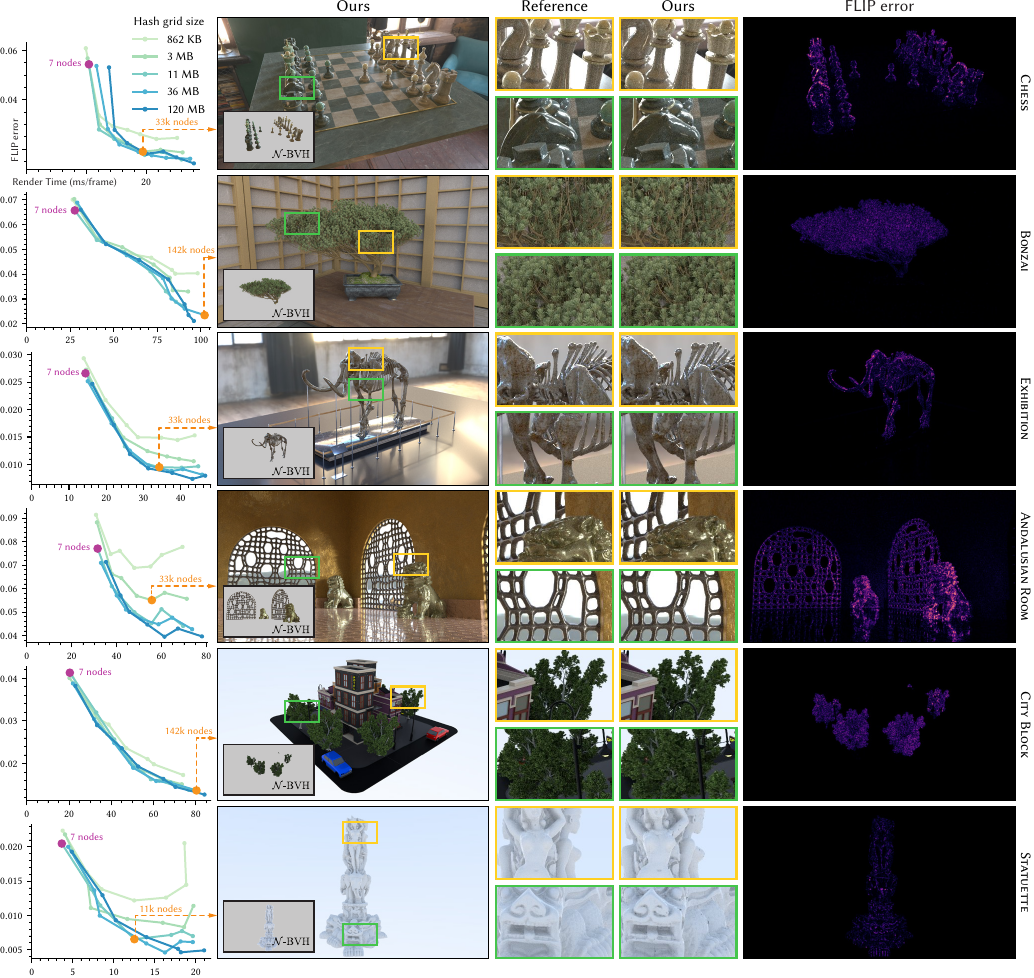}
        \captionof{figure}{
            Hybrid path tracing results, rendered at 1920$\times$1080 resolution. On the left, we plot the rendering time of our approach (which is a function of the \nbvh node count and hash-grid size) with respect to image error. Both the number of nodes in our \nbvh and the hash-grid size impact performance, the latter however less significantly. On the right, we show the rendered images and their FLIP error for a chosen configuration (\textcolor[RGB]{247, 148, 29}{\raisebox{-0.35mm}{\CIRCLE}}) that achieves a good performance/quality trade-off. In \cref{tab:performance_memory} we compare the rendering times and memory footprint of two configurations (\textcolor[RGB]{182, 62, 151}{\raisebox{-0.2mm}{\CIRCLE}}, \textcolor[RGB]{247, 148, 29}{\raisebox{-0.2mm}{\CIRCLE}}) to classical (i.e., non-neural) path tracing.
        }
        \label{fig:main_results}
    \end{center}
    \vspace{20mm}
\end{figure*}
%%%%%%%%%%%%%%%%%%%%%%%%%%%%%%%%%%%%%%%%%%%%%%%%%%%%%%%%%%%%

%%%%%%%%%%%%%%%%%%%%%%%%%%%%%%%%%%%%%%%%%%%%%%%%%%%%%%%%%%%%
\section{Implementation}
%%%%%%%%%%%%%%%%%%%%%%%%%%%%%%%%%%%%%%%%%%%%%%%%%%%%%%%%%%%%

We implemented our method in a \revision{fully software-based CUDA wavefront path tracer}. For training and inference, we use the \emph{tiny-cuda-nn} library \cite{Mueller22} with half-precision scalars. All our experiments are performed on an NVIDIA RTX 3090 GPU, except for neural prefiltering (\cref{sec:neural_prefiltering}), which is run on an NVIDIA RTX 3080 GPU to match the timings of \citet{neurallod2023}. \revision{Further implementation details and pseudo-code of our inference pipeline can be found in our supplemental document.}

%%%%%%%%%%%%%%%%%%%%%%%%%%%%%%
\paragraph{BVH construction}
%%%%%%%%%%%%%%%%%%%%%%%%%%%%%%

Since our \nbvh construction requires direct access to the base BVH, we cannot leverage hardware-accelerated construction and traversal. We construct that BVH on the CPU using a sweeping SAH builder \cite{spatial_split_2009}, upload it to the GPU, \revision{and traverse it in a dedicated CUDA kernel.}

%%%%%%%%%%%%%%%%%%%%%%%%%%%%%%
\paragraph{Neural model \& training}
%%%%%%%%%%%%%%%%%%%%%%%%%%%%%%

In all our experiments, we perform gradient descent in batches of $2^{18}$ rays and employ the Adam optimizer \cite{kingma2014adam} with default hyper-parameters and a learning rate of 0.01. The output MLP comprises 4 hidden layers, each containing 64 neurons with ReLU activations. The output layer has sigmoid activation for all ray-query outputs, except for the normal for which a linear activation has shown improved reconstruction quality. The hash grid contains 8 levels, \revision{starting from a base resolution of $8^3$ to a maximum resolution of $1024^3$}, with 4 features per level. To control the \revision{network's} memory footprint, we vary only the hash-map size. In scenes where the BVH nodes have extremely thin bounding boxes, training-time node intersection can be subject to floating-point errors. To alleviate this issue, we inflate \nbvh nodes slightly, following \citet{neurallod2023}.

% %%%%%%%%%%%%%%%%%%%%%%%%%%%%%%
% \paragraph{Depth estimation}
% %%%%%%%%%%%%%%%%%%%%%%%%%%%%%%

% \revision{
% The \emph{order} of the ray-query's concatenated features allows us to estimate depth from within nodes. The default order follows the ray direction; when a ray query is issued from within a node, we reverse that order. Then, an intersection occurs inside the node only if the inferred distance is smaller than the distance from the ray origin to the node's exit point.
% }

%%%%%%%%%%%%%%%%%%%%%%%%%%%%%%%%%%%%%%%%%%%%%%%%%%%%%%%%%%%%
\section{Application: Hybrid path tracing}
\label{sec:hybrid_path_tracing}
%%%%%%%%%%%%%%%%%%%%%%%%%%%%%%%%%%%%%%%%%%%%%%%%%%%%%%%%%%%%

The most direct use of our method is to replace traditional ray-tracing operations for large assets with our \nbvh, to reduce the overall memory footprint of the scene. We use a two-level hierarchy where the top-level acceleration structure (TLAS) holds several bottom-level structures (BLAS) at its leaves. In this hybrid hierarchy, each BLAS is either a classical BVH or our \nbvh. Whenever a leaf node in a BLAS is reached, a classical or neural ray intersection query is performed; both query types yield the same type of intersection data. The shading frame and BSDF are instantiated and sampled as usual to determine the next scattering direction for path tracing.

%%%%%%%%%%%%%%%%%%%%%%%%%%%%%%%%%%%%%%%%%%%%%%%%%%%%%%%%%%%%
\subsection{Results}
%%%%%%%%%%%%%%%%%%%%%%%%%%%%%%%%%%%%%%%%%%%%%%%%%%%%%%%%%%%%

In addition to the results presented next, our supplemental video demonstrates our real-time \nbvh construction and inference scheme. The results of our ablation and the renders for all our test scenes can be found in full in the supplemental document and HTML viewer.
\revision{To compare rendered images to their ground truths, we use the FLIP error metric~\cite{Andersson2020}.}

%%%%%%%%%%%%%%%%%%%%%%%%%%%%%%
\paragraph{Reconstruction quality \& performance}
%%%%%%%%%%%%%%%%%%%%%%%%%%%%%%

In \cref{fig:main_results} we plot the rendering error of various \nbvh configurations on six scenes. For each scene, we show renders and error maps for an \nbvh configuration that achieves a good trade-off between render time and reconstruction quality. Images for the remaining configurations can be found in the supplemental HTML viewer. These confirm that render times and reconstruction quality correlate most with the \nbvh node count, less so with the hash-grid size. Note that we deliberately select the most complex assets to be represented neurally; the rest of the scene does not benefit as much from high compression rates. In \cref{tab:performance_memory} we also compare performance and memory footprint to those of a traditional CUDA path tracer.

%%%%%%%%%%%%%%%%%%%%%%%%%%%%%%

\begin{figure}
    % \vspace*{80mm}
    \setlength{\tabcolsep}{2pt}
    \captionof{table}{
        Performance comparison against classical path tracing. Our approach achieves 2--4$\times$ higher render times. However, our representation delivers drastic memory compression at low error, allowing the rendering of complex scenes that could not even be uploaded onto low-end GPUs. Reducing the node count of our \nbvh (\textcolor[RGB]{182, 62, 151}{\raisebox{-0.3mm}{\CIRCLE}}) closes the gap to software path-tracing performance (albeit at a higher error), showing the flexibility of our approach to adapt to strict render-time budgets.
    }
    \vspace{-2.5mm}
    \scalebox{0.82}{
        \begin{tabularx}{1.208\linewidth}{l rr rrr rrr}
            \toprule
                                  & \multicolumn{2}{c}{\textbf{Path tracing}} & \multicolumn{3}{c}{\textbf{Our hybrid} \textcolor[RGB]{182, 62, 151}{\raisebox{-0.35mm}{\CIRCLE}}} & \multicolumn{3}{c}{\textbf{Our hybrid} \textcolor[RGB]{247, 148, 29}{\raisebox{-0.35mm}{\CIRCLE}}}                                                                                     \\
            \cmidrule(lr){2-3}
            \cmidrule(lr){4-6}
            \cmidrule(lr){7-9}
            \textbf{Scene}        & \textbf{Time}                             & \textbf{Memory}                                                                                    & \textbf{Time}                                                                                      & \textbf{Memory} & \textbf{FLIP} & \textbf{Time} & \textbf{Memory} & \textbf{FLIP} \\
            \midrule
            \textsc{Chess}        & 6.6\,ms                                   & 329\,\MB\;\;                                                                                       & 10\,ms                                                                                             & 36\,\MB\;\;     & 0.057         & 19\,ms        & 37\,\MB\;\;     & 0.019         \\
            \textsc{Bonzai}       & 28\,ms                                    & 853\,\MB\;\;                                                                                       & 29\,ms                                                                                             & 181\,\MB\;\;    & 0.069         & 102\,ms       & 186\,\MB\;\;    & 0.023         \\
            \textsc{Exhibition\;} & 13\,ms                                    & 2.69\,\GB\;\;                                                                                      & 14\,ms                                                                                             & 181\,\MB\;\;    & 0.027         & 34\,ms        & 182\,\MB\;\;    & 0.010         \\
            \textsc{And.\ Room}   & 28\,ms                                    & 309\,\MB\;\;                                                                                       & 31\,ms                                                                                             & 108\,\MB\;\;    & 0.088         & 55\,ms        & 109\,\MB\;\;    & 0.055         \\
            \textsc{City Block}   & 24\,ms                                    & 1.55\,\GB\;\;                                                                                      & 22\,ms                                                                                             & 180\,\MB\;\;    & 0.039         & 80\,ms        & 185\,\MB\;\;    & 0.014         \\
            \textsc{Statuette}    & 4.9\,ms                                   & 642\,\MB\;\;                                                                                       & 3.7\,ms                                                                                            & 10.8\,\MB\;\;   & 0.020         & 12\,ms        & 11.2\,\MB\;\;   & 0.007         \\

            \bottomrule
        \end{tabularx}
    }
    \label{tab:performance_memory}
\end{figure}
%%%%%%%%%%%%%%%%%%%%%%%%%%%%%%

%%%%%%%%%%%%%%%%%%%%%%%%%%%%%%
\paragraph{Compression rate}
%%%%%%%%%%%%%%%%%%%%%%%%%%%%%%

\begin{figure}
    \centering
    \includegraphics[width=\linewidth]{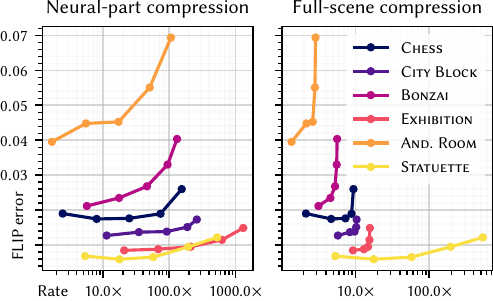}%
    \caption{
        Memory compression rates achieved on our test scenes. We keep the \nbvh node counts fixed to those chosen in \cref{fig:main_results} (\textcolor[RGB]{247, 148, 29}{\raisebox{-0.35mm}{\CIRCLE}}) and vary the hash-grid size along the horizontal axis. We plot the compression rates achieved on the neural part only and on the full scene.
    }
    \label{fig:compression_rates}
\end{figure}

The main strength of our approach lies in its ability to compress large and complex scenes into a few tens of megabytes, while maintaining their usability in a traditional rendering pipeline. In \cref{fig:compression_rates} we plot the different compression rates achieved on the scenes from \cref{fig:main_results}. We distinguish between the rate of the entire scene and the rate of only the portion represented by our neural model. On that portion alone we can achieve over $1000\times$ compression rates over the original geometry footprint. On the entire scene, the compression rates are $5-100\times$. The footprint reduction is mostly due to the geometry representation, though textured albedo and normal data are also learned by our model and contribute to the overall compression. Note that the mammoth in the \textsc{Exhibtion} scene has extremely high complexity as it is an original consolidated 3D scan from the Smithsonian Institute.

%%%%%%%%%%%%%%%%%%%%%%%%%%%%%%%%%%%%%%%%%%%%%%%%%%%%%%%%%%%%

%%%%%%%%%%%%%%%%%%%%%%%%%%%%%%%%%%%%%%%%%%%%%%%%%%%%%%%%%%%%

%%%%%%%%%%%%%%%%%%%%%%%%%%%%%%
\paragraph{Level of detail}
%%%%%%%%%%%%%%%%%%%%%%%%%%%%%%

A simple application of LoD already improves the performance of our hybrid path tracing: we switch to a coarser level in our \nbvh after the primary-ray intersection. Since the number of leaf nodes to query in the \nbvh can be drastically reduced, we achieve 1.5--2$\times$ faster rendering with only slight increase in reconstruction error. In the supplemental document we show additional results on the scene from \cref{fig:teaser} to demonstrate the effect on reconstruction quality and render time. Note that no other hybrid path tracing results in the paper use this LoD strategy. We do apply an LoD scheme in our neural prefiltering application (\cref{sec:neural_prefiltering}).

%%%%%%%%%%%%%%%%%%%%%%%%%%%%%%%%%%%%%%%%%%%%%%%%%%%%%%%%%%%%
\subsection{Ablation study}
\label{sec:ablation}
%%%%%%%%%%%%%%%%%%%%%%%%%%%%%%%%%%%%%%%%%%%%%%%%%%%%%%%%%%%%

We next evaluate how the hyperparameters of our model affect its size, training time, inference cost, and reconstruction quality.

%%%%%%%%%%%%%%%%%%%%%%%%%%%%%%
\paragraph{Hash-grid size vs.\ node count}
%%%%%%%%%%%%%%%%%%%%%%%%%%%%%%

\begin{figure}
    \begin{minipage}[t]{.27\textwidth}
        \centering
        \includegraphics[width= \textwidth]{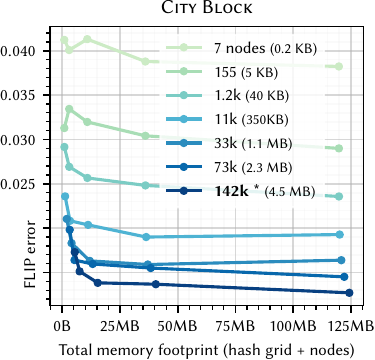}%
        \captionof{figure}{
            Total memory footprint of our representation vs.\ rendering error. Along each curve we vary the hash-grid size; that size impacts the error much less than the \nbvh node count (which is fixed along each curve and its footprint is reported in the legend). The asterisk indicates the node count chosen in \cref{fig:compression_rates,fig:main_results} (\textcolor[RGB]{247, 148, 29}{\raisebox{-0.25mm}{\CIRCLE}}).
        }
        \label{fig:memory_ablation}
    \end{minipage}%
    \hfill
    \begin{minipage}[t]{.185\textwidth}
        \centering
        \includegraphics[width=\textwidth]{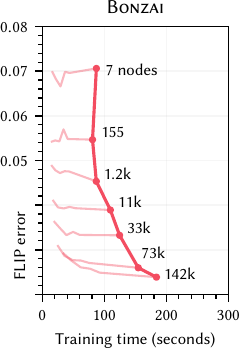}%
        \captionof{figure}{
            Training time vs.\ error. Hash-grid size is fixed. After an initial 1000-iteration cut optimization to reach a set total node count (indicated on plot), we plot error over the final 5000-iteration training-only stage.
        }
        \label{fig:train_time_ablation}
    \end{minipage}
\end{figure}

We observe that the biggest improvement in reconstruction quality is achieved by increasing the number of \nbvh nodes rather than the hash-grid size. Node count impacts memory footprint to a much lesser degree than the number of learnable hash-grid parameters. \Cref{fig:memory_ablation} illustrates how increasing the hash-grid size on the \textsc{City Block} scene has diminishing returns on reconstruction quality and rapidly plateaus as the necessary neural capacity is reached. All our test scenes exhibit similar behavior.

%%%%%%%%%%%%%%%%%%%%%%%%%%%%%%

%%%%%%%%%%%%%%%%%%%%%%%%%%%%%%

%%%%%%%%%%%%%%%%%%%%%%%%%%%%%%
\paragraph{Training time}
%%%%%%%%%%%%%%%%%%%%%%%%%%%%%%

Training our entire pipeline takes only a couple of minutes. We observe that long training times are not necessary to achieve good reconstruction quality. What impacts training time (and error) the most is again the \nbvh node count, since a larger node count increases the traversal time. In \cref{fig:train_time_ablation} we fix the hash-grid size and plot the training times achieved for different node counts on the \textsc{Bonzai} scene, our geometrically most complex scene. The same experiment on all our test scenes, with similar results, can be found in the supplemental document.

%%%%%%%%%%%%%%%%%%%%%%%%%%%%%%
\paragraph{Hash-grid utilization}
%%%%%%%%%%%%%%%%%%%%%%%%%%%%%%

While the leaves of our \nbvh occupy the 3D space covered by the hash-grid model only very sparsely, the model's \emph{capacity} is still highly utilized thanks to the hash collisions naturally dispersing encoded queries across its learnable features. The supplemental document contains a more detailed table comparing the volume occupied by the \nbvh leaves on our test scenes and the average number of non-zero gradients during training for different hash-map sizes.

%%%%%%%%%%%%%%%%%%%%%%%%%%%%%%%%%%%%%%%%%%%%%%%%%%%%%%%%%%%%
\section{Application: Neural prefiltering}
\label{sec:neural_prefiltering}
%%%%%%%%%%%%%%%%%%%%%%%%%%%%%%%%%%%%%%%%%%%%%%%%%%%%%%%%%%%%

\citet{neurallod2023} showed that neural appearance prefiltering can save large amounts of memory while maintaining low perceptual error. Their method traverses a neural sparse-voxel visibility-only representation to find the first voxel that reports occlusion along the ray. Their appearance sparse-voxel network then predicts the prefiltered appearance (represented as a phase function) for a randomly sampled direction along which the path is continued. We plug into their pipeline, substituting their visibility grid with our \nbvh. We keep their appearance network, directly evaluating their publicly available pre-trained weights.

%%%%%%%%%%%%%%%%%%%%%%%%%%%%%%
\paragraph{Improved rendering performance}
%%%%%%%%%%%%%%%%%%%%%%%%%%%%%%

The neural visibility grid of \citet{neurallod2023} incurs a major performance overhead as it performs inference in each of potentially many traversed voxels in search for an intersection. Note that they do not require explicit intersection-distance estimation as their appearance network prefilters the occlusion \emph{inside a voxel} (at a given level of detail).

By replacing their grid with our \nbvh, we drastically reduce the number of required visibility inferences. This is due to a single \nbvh leaf typically covering much larger space than one of their voxels. Once an intersection is found using our \nbvh, we only need to identify in which appearance voxel the intersection took place. We then carefully take into account their voxel-based path-space formulation to trace secondary rays, which requires the ray origin to lie on the previously determined appearance-voxel's boundary rather than on the intersected surface.

\begin{figure}
    \includegraphics[width=\columnwidth]{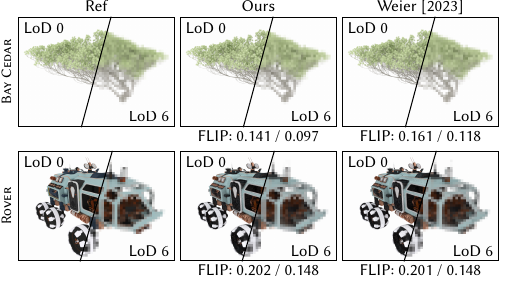}\\
    \vspace{-2mm}
    \captionof{figure}{
        Neural appearance prefiltering comparison. Our \nbvh method achieves equal or better reconstruction quality than that of \citet{neurallod2023} on both structured and unstructured geometry at all scales, with faster rendering and training times, at lower memory footprint (see \cref{fig:neural_prefiltering_stats}).
    }
    \label{fig:vs_prefiltering}
    \vspace{-2mm}
\end{figure}

\begin{figure}
    \centering
    \begin{minipage}{.6\linewidth}
        \setlength{\tabcolsep}{0.5pt}
        \scalebox{0.90}{
            \begin{tabularx}{1.047\linewidth}{lcc}
                \toprule
                \textbf{Scene}    & \textbf{Training}     & \textbf{~Network\,$+$\,Accel.} \\
                \midrule
                \textsc{Cedar}    &                       &                                \\
                \;\;\textbf{Ours} & \textbf{~\,3:25\,min} & \textbf{10.8 $+$ 5.7 \MB}      \\
                \;\;[Weier\,'23]  & 10:57\,min            & 30.1 $+$ 8.9 \MB               \\[1mm]
                \midrule
                \textsc{Rover}    &                       &                                \\
                \;\;\textbf{Ours} & \textbf{2:11\,min}    & \textbf{10.8 $+$ 0.7 \MB}      \\
                \;\;[Weier\,'23]  & 6:14\,min             & 15.1 $+$ 9.5 \MB               \\
                \bottomrule
            \end{tabularx}
        }
    \end{minipage}%
    \begin{minipage}{0.4\linewidth}
        \includegraphics[width=0.99\linewidth]{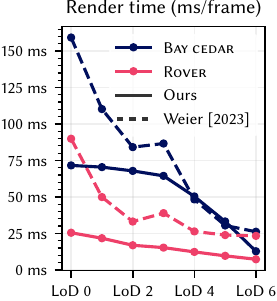}
    \end{minipage}
    \vspace{-2.5mm}
    \captionof{figure}{
        Training-time and memory-footprint comparison of our appearance prefiltering method against that of \citet{neurallod2023}, \revision{at LoD 0 (highest resolution)}. They use a voxel grid as a ray-acceleration structure, while we use our \nbvh. The timing and memory numbers exclude the appearance network. Rendering times are reported for $1024^2$ resolution.
    }
    \label{fig:neural_prefiltering_stats}
    % \vspace{-1mm}
\end{figure}

In \cref{fig:vs_prefiltering}, we show that our approach achieves similar reconstruction quality on both structured and unstructured geometry; the remaining error is mostly due to the inaccuracies in the appearance network. We matched the seven appearance LoDs with visibility LoDs in our \nbvh using as many tree cuts. For the \textsc{Bay Cedar} scene, our \nbvh contains 127 nodes at the coarsest level and 176k nodes at the finest. The \textsc{Rover} scene requires a much lower node count, 91 and 21k respectively. Hash-map size is $2^{18}$ for both scenes. On the right of \cref{fig:neural_prefiltering_stats} we report the render times and compare them to the authors' implementation. Theirs includes a visibility Russian-roulette mechanism to gain a $2-3\times$ speed up over their vanilla implementation. Our approach does not require such a scheme and still easily yields a 2$\times$ speed improvement.

%%%%%%%%%%%%%%%%%%%%%%%%%%%%%%
\paragraph{Memory footprint \& training}
%%%%%%%%%%%%%%%%%%%%%%%%%%%%%%

In \cref{fig:neural_prefiltering_stats}, we compare the memory footprint and training times of the two approaches. Our \nbvh exhibits much faster training and lower memory footprint as we can aggressively reduce the hash-grid size without loss of quality.

%%%%%%%%%%%%%%%%%%%%%%%%%%%%%%
%%%%%%%%%%%%%%%%%%%%%%%%%%%%%%

%%%%%%%%%%%%%%%%%%%%%%%%%%%%%%
%%%%%%%%%%%%%%%%%%%%%%%%%%%%%%

%%%%%%%%%%%%%%%%%%%%%%%%%%%%%%%%%%%%%%%%%%%%%%%%%%%%%%%%%%%%
\section{Discussion}
%%%%%%%%%%%%%%%%%%%%%%%%%%%%%%%%%%%%%%%%%%%%%%%%%%%%%%%%%%%%

Our \nbvh delivers intersection distance, normal, and appearance attributes for ray queries. It integrates by design into existing path-tracing pipelines, and offers control over intuitive trade-offs (hierarchy depth, neural capacity) to fit various application scenarios. Its performance is output-sensitive, depending on the amount of allocated storage, the frequency of the scene's content, and its self-similarity, rather than on the number of input primitives.

%%%%%%%%%%%%%%%%%%%%%%%%%%%%%%
\paragraph{Limitations}
%%%%%%%%%%%%%%%%%%%%%%%%%%%%%%

The main limitation in our neural ray query model lies in the assumption of convexity (or concavity) of the geometry inside a neural node to ensure correct intersection estimation when querying the model with a ray originating within the node's bounds. This can lead to incorrect offsets of secondary ray origins when using a low number of \nbvh nodes; as the tree grows deeper, the problem disappears since convexity/concavity is ensured in the limit. In practice, for the scenes we tested, even at a low node count (1.2k), this issue did not manifest. One existing constraint in our encoding scheme is the fixed number of encoded points per node. Although we have identified that three points strike a good balance between inference speed and reconstruction quality, introducing an adaptive sampling approach---increasing points in challenging areas and reducing them in simpler regions---could allow for a more balanced workload between traversal cost and inference time.

%%%%%%%%%%%%%%%%%%%%%%%%%%%%%%
\paragraph{Future work}
%%%%%%%%%%%%%%%%%%%%%%%%%%%%%%

Our approach can be further generalized. Since we fit our representation solely based on the result of ray queries, it is readily applicable to any surface representation that can be intersected. Being based on a standard BVH, it can also inherit future improvements brought to this primitive partitioning structure. \revision{Our current method prevents compression of \emph{dynamic} geometry since a naïve solution would require to re-train our representation at every frame. Further extending our neural ray-query encoding to handle dynamic content would be an interesting direction for future work.}

%%%%%%%%%%%%%%%%%%%%%%%%%%%%%%
\paragraph{Conclusion}
%%%%%%%%%%%%%%%%%%%%%%%%%%%%%%

\begin{figure}
    \centering
    \includegraphics[width=1\linewidth]{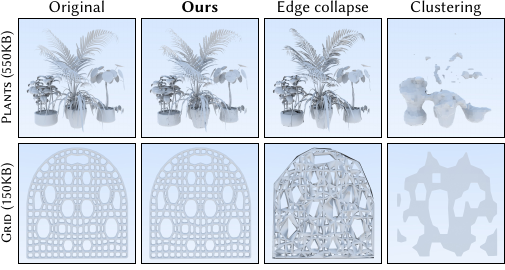}%
    \vspace{-2mm}
    \captionof{figure}{
        Geometric simplification comparison. We apply classical edge collapse \cite{garland97} and clustering \cite{Schaefer:2003:AVC} to reduce the memory footprint of two scenes from 42\,MB to 550\,KB and from 9\,MB to 150\,KB respectively. We compare these to our approach with equal footprint. Our method is more robust to the diversity of signals found in a complex 3D scene than these specialized surface- or spatial-clustering approaches.
    }
    \label{fig:vs_decimation}
    \vspace{-1mm}
\end{figure}

\nbvh drastically compresses complex scenes while maintaining low rendering error, offering graceful degradation as capacity decreases and blending trivially into classical pipelines. Designed to be efficiently queried by rays, our approach can effectively compress the geometric representation much more accurately than standard geometric simplification techniques such as \emph{edge-collapse} or \emph{spatial clustering} (\cref{fig:vs_decimation}). Although we cannot yet compete with the speed of raw path-tracing algorithms, we enable the rendering of scenes that would not otherwise fit in memory, and we think that porting our \nbvh on the hardware could benefit from fused neural and hardware ray-tracing operations, i.e., the ability to evaluate our neural model on the fly while traversing the tree, avoiding heavy context switches for each inference.

%%%%%%%%%%%%%%%%%%%%%%%%%%%%%%
%%%%%%%%%%%%%%%%%%%%%%%%%%%%%%

%%%%%%%%%%%%%%%%%%%%%%%%%%%%%%%%%%%%%%%%%%%%%%%%%%%%%%%%%%%%
% Acknowledgments
%%%%%%%%%%%%%%%%%%%%%%%%%%%%%%%%%%%%%%%%%%%%%%%%%%%%%%%%%%%%

\begin{acks}
    We thank Jean-Marc Thiery and Lo\"{i}s Paulin, Pascal Grittmann, \"{O}mercan Yazici, and the Adobe Research team for sharing insights throughout the project. We also thank the anonymous reviewers for their feedback and suggestions. We have used assets from diverse artists: PolyHaven (environment maps), Walt Disney Animation Studio (\textsc{Bay Cedar} from the Moana Island Scene) and vajrablue (\textsc{Rover}). This project has received funding from the \grantsponsor{PRIME}{European Union’s Horizon 2020 research and innovation programme}{https://rea.ec.europa.eu/funding-and-grants/horizon-europe-marie-sklodowska-curie-actions_en} under the \href{https://prime-itn.eu}{Marie Sk\l{}odowska-Curie} grant agreement No\,956585.
\end{acks}

%%%%%%%%%%%%%%%%%%%%%%%%%%%%%%%%%%%%%%%%%%%%%%%%%%%%%%%%%%%%
% Bibliography
%%%%%%%%%%%%%%%%%%%%%%%%%%%%%%%%%%%%%%%%%%%%%%%%%%%%%%%%%%%%

\bibliographystyle{ACM-Reference-Format}
\bibliography{bibliography}

\end{document}